\documentclass[12pt]{article}
\def\dgr{\dagger}
\def\beq{\begin{equation}}
\def\eeq{\end{equation}}
\def\ba{\begin{eqnarray}}
\def\ea{\end{eqnarray}}

\begin{document}
\renewcommand{\thefootnote}{\fnsymbol{footnote}}
%\begin{titlepage}
\begin{flushright}
\begin{tabular}{l}
UPR-919-T\\
hep-ph/0101002
\end{tabular}
\end{flushright}
\vskip0.5cm
\begin{center}
{\LARGE\bf
CP Violation and Extra Dimensions}

\vspace*{0.8cm}
{\large\bf
Chao-Shang Huang$^1$, Tianjun Li$^2$, 
Liao Wei$^1$, Qi-Shu Yan$^3$} 
\vglue 0.5cm
%\vspace*{0.2cm}
\begin{flushleft}
$^1$ITP, Academia Sinica, P. O. Box
2735, Beijing 100080, P. R. China \\
$^2$Department of Physics and Astronomy, University of Pennsylvania,
Philadelphia, PA 19104-6396, U. S. A.\\
$^3$ Physics Department of Tsinghua University,
             P. R. China \\
\end{flushleft}
\vglue 0.2cm
      
\bigskip

%{\em Version of \today}

%\vspace{1cm}

\bigskip

{\large\bf Abstract\\[10pt]}

\end{center}
It is shown that the new sources of CP violation can be generated in the 
models with more than one extra dimensions. In the supersymmetric models
on 
the space-time $M^4\times T^2/Z_2$, where the radius moduli have auxiliary 
vacuum expectation values and the supersymmetry breaking is mediated by 
the Kaluza-Klein states of gauge supermultiplets, we analyze the gaugino
masses and trilinear couplings for two scenarios and obtain that 
there exist relative CP violating phases among the gaugino masses and 
trilinear couplings.

\bigskip

\centerline{{\sc Pacs} numbers: 11.30.E, 04.50, 12.60, 11.30.Q,}
%\end{titlepage}
\newpage
\renewcommand{\thefootnote}{\arabic{footnote}}
\setcounter{footnote}{0}
%\section{\bf Introduction}
It is well-known that the consistent weakly coupled
(perturbative) superstring theories exist 
only in ten dimensions because of anomaly cancellations
and the extra six dimensions must be compactified so that 
the universe we ``see"
is four-dimensional. In the weakly coupled heterotic
$E_8\times E_8$ string, which was
the most phenomenologically interesting candidate 
among the known perturbative superstring theories, 
the compactification energy scale is the unification scale 
$M_{GUT}\sim 2\times 10^{16} GeV$,
which is too high to be probed.
 Recently, the study of the strongly coupled superstring
theories and duality opens a way which leads to the possibility to have
large extra
dimension compactifications due to the presence of the brane. 
For example, in the type I/I' string theory \cite{11} the six compact
dimensions
are separated into tangent and transverse to the D-branes and the string
scale can be
made much smaller (e.g., 1 TeV) than the Planck scale if the physical
volume of
the transverse dimensions are very large \cite{ant}. 
Consistent with the compactification picture of string theories, various
models with large extra dimensions have been proposed and their
phenomenological implications
in particle physics, gravity and astrophysics have extensively been
examined \cite{nath} since
the pioneered works \cite{add} were published.

On the other hand, the origin of CP violation has been one of main issues
in high energy physics since the discovery of CP violation in 
the $K_0-\overline{K}_0$ system in 1964 \cite{new1}.
The observation of Re($\epsilon'$/$\epsilon$) by KTeV collaboration
\cite{kt} definitely confirms the earlier NA31 experiment~\cite{na}. This
direct CP violation measurement in the kaon system can be accommodated by
the CKM phase in the Standard Model within theoretical uncertainties.
Recently the results on CP violation in 
$B_d$ - $\bar B_d$ mixing have been reported by the BaBar and Belle
Collaborations 
\cite{Osaka} in the ICHEP2000 Conference, which can also be explained in 
the Standard Model within
 both theoretical and experimental uncertainties.
However, the CKM phase is not enough to explain the matter-antimatter
asymmetry 
in the universe and gives the contribution to 
the electric dipole moments (EDMs) of the neutron and 
electron much smaller than the experimental bounds 
of the EDMs of electron and neutron. One needs to
have new sources of CP violation in addition
to the CP violation from CKM matrix,  which has been one 
of the motivations to search new theoretical models beyond the 
Standard Model, and examine their phenomenological effects.

Although vast phenomenological implications have been studied in
various models with large
extra dimensions, the CP violation\footnote{There are papers 
in which the CP violation is discussed in the models with extra dimensions
\cite{sak}. However, in their models, the origin of CP violation is of
complex vacuum expectation
values of Higgs fields, which is not directly related to extra
dimensions.}
has not been 
examined so far. In this letter, we shall show that it is possible to have
a new source of CP violation due to the presence of extra dimensions. 
Assuming the 6-dimensional
space-time manifold is $M^4\times T^2/Z_2$, we calculate the gaugino
masses and trilinear couplings for two scenarios in a framework
where the radius moduli, which are related to the physical sizes
of the extra dimensions, have auxiliary vacuum expectation values, and
the supersymmetry breaking is mediated by the Kaluza-Klein states
of gauge supermultiplets. We obtain that 
there are relative CP violating phases among the gauigno masses and
trilinear couplings. Furthemore, it is easy to generalize our scenarios to
the supersymmetric models with more than two large extra dimensions and
the CP violation can be induced by similar mechanism.
Therefore, we conclude that the new sources of CP violation can be
generated in the
supersymmetric models with more than one large extra dimensions
where radius moduli have auxiliary vacuum expectation values and
the supersymmetry breaking is mediated by the Kaluza-Klein states
of gauge supermultiplets. Note that
in 5-dimensional SUSY theories, there is no non-trivial CP violation
induced
due to the SUSY breaking because the overall phase can be rotated 
away~\cite{cl,ky}.

%\section{\bf Soft terms induced by two radions}

For our purpose, we consider a N=1 6-dimensional supersymmetric
(SUSY) theory compactified 
on $T^2/Z_2$. Upon 
compactification, in the 4-dimensional effective theory
 there are two modulus superfields, $X_1$ and $X_2$, which are related to
the 
 radii $R_1$ and $R_2$ of the torus. We assume that the observable sector
is located at one of the four different fixed points of the orbifold
$T^2/Z_2$ and the modulus
superfields have auxiliary vacuum expectation values, 
$\langle X_a \rangle = M_a + \theta^2 F_a$,
where a=1,2 and $M_a \sim R_a^{-1}$.
We shall consider two scenarios: (A) the Standard Model gauge fields all
propagate in
the bulk; (B) the SU(3) and SU(2) $\times U(1)$ propagate 
in different extra dimensions.  
In scenario A, we obtain that there are relative CP violating
phases between the gaugino masses and the trilinear
couplings, and among the trilinear couplings. In scenario B,
we also have different
phases among the gaugino masses, in addition to those in scenario A. 

As pointed out in Ref.~\cite{ls}, before Weyl-rescaling there are no
(non-derivative) direct 
couplings of the modulus $X_a$ to the observable sector.  The reason is
that $X_a$ can only couple to 
the higher-dimensional components of the energy-momentum tensor and the
wave functions of KK 
zero-mode fields do not depend on the extra-dimensional coordinates. The
couplings of 
the KK (non-zero) modes of gauge supermultiplets to modulus fields will
give the mass 
splitting in the SUSY 
multiplets. Thus, the KK excitations of gauge supermultiplets act as
messenger fields which 
transmit the SUSY breaking effect to the 
observable sector and consequently the soft terms are 
generated at a quantum level in the four-dimensional effective theory. We
shall use the method
given in Ref.~\cite{extract}, i.e., from wave function renormalization, to
derive the soft terms. 

First, we consider the scenario A. Without loss
of generality, we assume  $M_2 < M_1$.
The messenger mass spectra can be derived from the couplings
of KK excitations of gauge supermultiplets to the background superfields 
\beq
\int d\theta^2d\bar\theta^2 \sum_{a=1}^2 
    X^+_a e^{n_a V^{n_1,n_2}} X_a ~,~~~~~~~~~~~n_1,n_2=0,1,2,...
\eeq
where $V^{n_1,n_2}$ is the KK mode with mass
\beq
m^2_{n_1,n_2}= \sum_{i=1}^2 \frac{n_i^2}{R_i^2}~.
\eeq
The gaugino masses are given by
\beq
{\tilde M}_i (\mu )=\left. -\frac{1}{2}\sum_{a=1,2} \frac{\partial \ln
S_i(X_a,\mu)}
{\partial \ln X_a}\right|_{X_a=M_a}~\frac{F_a}{M_a}~, 
\label{gauginoc}
\eeq
where $i= 1, 2, 3,$ which correspond
 to the Standard Model gauge groups U(1), SU(2) and SU(3), respectively.

Taking into account the contributions of KK excitations propagating in
extra dimensions, we express the running gauge couplings as 
\ba
     \alpha_i^{-1}(\mu) ~=~
      \alpha_i^{-1}(\Lambda)
  ~-~ {(b_i-{\tilde b_i}) \over 2\pi} \,\ln{\mu \over \Lambda}
 ~-~ {\tilde b_i \over 4} \,\left\lbrack {\mu^2 \over {M_1 M_2}}
-{\Lambda^2 \over {M_1 M_2}} \right\rbrack~, {\rm ~for~} M_1 < \mu \leq
\Lambda~,
\label{strconi}
 \\
\alpha_i^{-1}(\mu) ~=~ \alpha_i^{-1}(M_1) 
~-~ {(b_i-{\tilde b_i}) \over 2\pi} \,\ln{\mu \over M_1} 
~-~ {\tilde b_i \over \pi} \,\left\lbrack {\mu \over M_2}
-{M_1 \over M_2} \right\rbrack~, {\rm ~for~} M_2 < \mu \leq M_1~,
\label{strcon}
\ea
where $\Lambda$ is the cutoff scale
 of the effective theory, $b_a$ and $\tilde{b}_a$ are 
the one-loop beta function coefficients in the MSSM and those
arising from the massive gauge boson multiplets~\cite{ddg}, respectively. 

From Eqs.~(\ref{gauginoc}), (\ref{strconi}) and (\ref{strcon}),
 it is straightforward to derive 
the gaugino masses 
\beq
{\tilde M}_i (\mu)=\frac{\alpha_i(\mu)}{4\pi} {\tilde b}_i \bigg [ 
\frac{N(\Lambda)}{2}(\frac{F_1}{M_1}+\frac{F_2}{M_2})
+(\frac{\pi}{2}-2)\frac{M_1}{M_2}(\frac{F_1}{M_1}-\frac{F_2}{M_2}) 
-\frac{F_2}{M_2} \bigg]~, 
\label{gaugino}
\eeq
where $ \mu < M_2$ and 
\ba
N(\mu)\simeq \left\{ \begin{array}{ll}
\pi \mu^2/M_1 M_2,  & ~~~~{\rm for}~ M_1 < \mu \leq \Lambda~, \\ 
2 \mu/M_2,  &~~~~{\rm for}~ M_2 < \mu \leq M_1~, \end{array} \right.
\ea
which is proportional to the surface
of an ellipse with long axis $R_2$ and short axis $R_1$,
indicating the number
of the KK excitations. 

Trilinear terms, $A_Q$, can be derived from the wave-function
renormalization
of chiral superfield Q, and can be expressed as
\ba
A_Q(\mu)&=&\sum_{a=1,2}\left . \frac{\partial \ln Z_Q(X_a,X_a^{\dgr},\mu)}
{\partial \ln X_a} \right|_{X_a=M_a} ~\frac{F_a}{M_a}~,
\ea
where $Z_Q$ can be solved out from the differential equations
\ba
\frac{d}{dt} \ln Z_Q&=&\sum_{j=1,2,3}\frac{C_Q(j)}{2\pi}
\tilde{\gamma}_Q^j(\mu)~,
\label{zq}
\ea
where
\ba
\tilde{\gamma}_Q^j(\mu)&=&
\left\{ \begin{array}{ll}
 \alpha_j(\mu), & ~~~~{\rm for}~  \mu < M_2~, \\
\alpha_j(\mu) N(\mu), &~~~~{\rm for}~  M_2 < \mu < \Lambda~, \end{array}
\right.  
\ea
where $C_Q(1), C_Q(2), C_Q(3)$ are the quadratic Casimirs of the Q 
representations of U(1), SU(2), SU(3), respectively (
$C_Q(j)=(j^2-1)/(2j)$ for
an SU(j) fundamental.).

To simplify the discussion, we only illustrate the contribution of one
gauge interaction, i.e., one term in the sum  in Eq. (\ref{zq}).
Then we have
\ba
A_Q^i(\mu)&=&\frac{C_{qi}}{4 \pi}\bigg\{(\alpha_i(M_2)(1-\frac{\tilde b_i}
{b_i})+\frac{\tilde b_i}{b_i}\alpha_i(\mu))
\bigg[\frac{N(\Lambda)}{2}(\frac{F_1}{M_1}+\frac{F_2}{M_2})
+(\frac{\pi}{2}-2)\frac{M_1}{M_2}(\frac{F_1}{M_1}-\frac{F_2}{M_2})
-\frac{F_2}{M_2}\bigg] \nonumber \\
&&-(\frac{b_i}{\tilde b_i}-1)\bigg[\frac{1}{2 D^i}
-\alpha_i(M_2)\bigg]\frac{F_1}{M_1} \bigg\}~,
\ea
where
\ba
\bar{D^i}=\alpha^{-1}(\Lambda)+\frac{\tilde
b_i}{4}\frac{\Lambda^2}{M_1M_2}
-\frac{b_i-{\tilde b_i}}{4\pi}ln\frac{M_1 M_2}{\Lambda^2}
+\frac{b_i-{\tilde b_i}}{4 \pi}~. 
\ea

For the case $M_1=O(M_2)$, where the running between $M_2$ and $M_1$ can
be neglected, the gaugino masses can be approximated to be 
\beq
{\tilde M_i}(\mu)=\frac{\alpha_i(\mu)}{4\pi}{\tilde
b_i}\bigg[\frac{N(\Lambda)}
{2}(\frac{F_1}{M_1}+\frac{F_2}{M_2})-{\pi \over 2}\frac{M_2}{M_1}
(\frac{F_1}{M_1}-\frac{F_2}{M_2}) -\frac{F_2}{M_2}\bigg]~,
\eeq
and the trilinear term reduces to
\ba
A_Q^i(\mu) &\doteq&
{C_{qi} \over b_i}\bigg[[1+\frac{\alpha_i(M_2)}{\alpha_i(\mu)}
(\frac{b_i}{\tilde b_i}-1)]{\tilde M_i(\mu)}+ (\frac{b_i}{\tilde b_i}-1)
\frac{\alpha_i(M_2)}{8 \pi} b_i
\frac{F_2}{M_2} \bigg]
\label{AT}~.
\ea
So, the gaugino masses and 
 trilinear terms will have different CP violating phases when
$N(\Lambda)$ is not much larger than one so that the last term of
Eq.~(\ref{AT})
could compete with the first one. This implies that the CP violating
effects is non-trivial. In addition, we notice that 
among the trilinear terms, the relative CP violating phases can also be
generated
for the up-type quark, down-type quark, and lepton have different charges 
under the Standard Model gauge groups.
Therefore, we conclude that the quantum effects can induce non-trivial 
CP violating phases.

Now we turn to the scenario B, in which the SU(3) and
SU(2)$\times$U(1) gauge superfields
 live in different extra dimensions.
 The calculations are straightforward and we only give
the results here. The gaugino masses are
\beq
{\tilde M}_3 (M_1)=\frac{\alpha_3(M_1)}{4\pi} {\tilde
b_3}N_1(\Lambda) \frac{F_1}{M_1}~,
\label{gau3}
\eeq
\beq
{\tilde M}_{1,2} (M_2)=\frac{\alpha_{1,2}(M_2)}{4\pi} {\tilde
b_{1,2}}N_2(\Lambda) \frac{F_2}{M_2}~,
\label{gau12}
\eeq
where $N_{1,2}(\mu)\simeq 2 (\mu/M_{1,2})$, indicating the number
of excited KK modes.

As for trilinear couplings $A_Q$, we obtain
\ba
A_D&=&
\sum_{q=D,Q}\alpha_3(\mu)\frac{C_{q3}}{2\pi} N_1(\Lambda) \frac{F_1}{M_1}
+ \sum_{q=D,Q} \sum_{i=1,2}\alpha_i(\mu)\frac{C_{qi}}{2\pi}
N_2(\Lambda) \frac{F_2}{M_2}
~,\label{eq:ad}\\
A_U&=&
\sum_{q=U,Q}\alpha_3(\mu)\frac{C_{q3}}{2\pi} N_1(\Lambda) \frac{F_1}{M_1}
+ \sum_{q=U,Q} \sum_{i=1,2}\alpha_i(\mu)\frac{C_{qi}}{2\pi}
N_2(\Lambda) \frac{F_2}{M_2}
~,\label{eq:au}\\
A_E&=&
 \sum_{q=E,L} \sum_{i=1,2} \alpha_i(\mu)\frac{C_{qi}}{2\pi}
N_2(\Lambda) \frac{F_2}{M_2}
~,\label{eq:ae}
\ea
where $C_{U2}=C_{D2}=0$.

If $F_1$ and $F_2$ had different phases, we obtain that
the phase and magnitude of $A_E$ are different from those of
$A_D$ and $A_U$ from Eqs. (\ref{eq:ad}-\ref{eq:ae}), and
the phase and magnitude of ${\tilde M}_3$ are different from those of 
${\tilde M}_1$ and ${\tilde M}_2$ from Eqs.~(\ref{gau3}) and
(\ref{gau12}).

In summary, we have shown by two specific scenarios that the
new sources of CP violation can be generated in the supersymmetric
models on the space-time $M^4\times T^2/Z_2$.
In general, provided that the number of extra dimensions is larger than
one, there are new sources of CP violation in the supersymmetric models
where the radius moduli have auxiliary vacuum expectation values and
the supersymmetry breaking is mediated by the Kaluza-Klein modes
of gauge supermultiplets. It should be pointed out that in the general
supersymmetric theories, although in many cases the sizes of CP violating
phases are strongly constrained by the experimental bounds on the electric
dipole moments
(EDMs) of the electron, neutron and $^{199}$Hg atom, the possible
cancellations
among the different contributions to the EDMs can significantly weaken the
upper bounds on the phases, therefore, the CP violating phases in the
supersymmetric theories can be large~\cite{cm}.

Another possibility to have CP violation is to construct the model
 in which the charge conjugation is conserved but the parity symmetry may
be
broken using
extra dimensions\cite{mp}. Of course, it is interesting to search 
other new sources of CP violation in the models with large extra
dimensions. 

Note added: after finishing the work, we noticed the e-preprint by Branco
et al\cite{bra}, in which the
CP violation in quark sector in AS scenario is discussed. 

\section*{\bf Acknowledgments}
The work was supported in part by the Natural Science
Foundation of China and by the U.S.~Department of Energy under Grant 
 No.~DOE-EY-76-02-3071.
%\section{\bf Appendix}

\end{document}